\documentclass[letterpaper, 10 pt, conference]{ieeeconf}
\IEEEoverridecommandlockouts

\usepackage{wrapfig}
\usepackage{threeparttable}% tables with footnotes
\usepackage{dcolumn}% decimal-aligned tabular math columns
\usepackage{nomencl}% automatic nomenclature generation via makeindex
\usepackage{subfigmat}% matrices of similar subfigures, aka small mulitples

\usepackage{cite}
\usepackage{graphicx}
\usepackage{epstopdf}
\usepackage{amsmath,amssymb,amsfonts,amsbsy,mathrsfs}
\usepackage[usenames, dvipsnames]{color}
\usepackage{caption}
\usepackage{url}
\usepackage{float}
\usepackage{listings}
\usepackage{longtable}
\usepackage{supertabular}
\usepackage{algorithmic}
\usepackage{algorithm}
\usepackage{mdframed}
\usepackage{mathabx}
\usepackage{xcolor}
\usepackage{amssymb}
\usepackage{enumerate}

\newtheorem{lemma}{Lemma}

\newtheorem{theorem}{Theorem}

\newtheorem{remark}{Remark}
\newtheorem{assumption}{Assumption}

\makeatletter
\newcommand{\To}[2][]{\ext@arrow 0359\Rightarrowfill@{#1}{#2}}
\makeatother

\DeclareMathAlphabet{\mathpzc}{OT1}{pzc}{m}{it}
\begin{document}

\title{\LARGE \bf Closeness of Solutions for Singularly Perturbed Systems via Averaging}

\author{Mohammad Deghat, Saeed Ahmadizadeh, Dragan Ne\v si\'c and Chris Manzie
\thanks{This work was supported by the ARC Discovery Scheme, grant number DP170104102.}
\thanks{The authors are with the Department of Electrical and Electronic Engineering, The University of Melbourne, Parkville, 3010, Victoria, Australia
{\tt\small \{m.deghat;ahmadizadeh.s;dnesic;manziec\}@unimelb}
{\tt\small .edu.au}}
}

\maketitle
\thispagestyle{empty}
\pagestyle{empty}
\allowdisplaybreaks
%=====================================
\begin{abstract}
This paper studies the behavior of singularly perturbed nonlinear differential equations with boundary-layer solutions that do not necessarily converge to an equilibrium. Using the average of the fast variable and assuming the boundary layer solutions converge to a bounded set, results on the closeness of solutions of the singularly perturbed system to the solutions of the reduced average and boundary layer systems over a finite time interval are presented. The closeness of solutions error is shown to be of order $\mathcal O(\sqrt\varepsilon)$, where $\varepsilon$ is the perturbation parameter.
%Moreover, under the additional assumption of exponential stability of the reduced average system, (semiglobal) practical exponential stability of the solutions of the singularly perturbed system is provided over the infinite time interval.
\end{abstract}
%=====================================
\section{Introduction}
The singular perturbation method is a common technique to analyze a two-time scale system via the behavior of two auxiliary systems, namely the reduced (slow) system and the boundary layer (fast) system. In general, the results using the singular perturbation method either relate the stability properties of the original system with the above-mentioned auxiliary systems or estimate the closeness of solutions of the original system to the solutions of the auxiliary systems; see e.g. \cite{kokotovic1999singular}, \cite[Sec. 11]{khalil2002nonlinear} for results on stability and closeness of solutions of the classical singular perturbation problem.
It is usually assumed in the classical singular perturbation results that the solutions of the boundary layer system converge to a unique equilibrium manifold. The case where the solutions converge to a bounded set, e.g. a set of limit cycles, has been studied using the averaging method \cite{artstein1998stability, grammel2008robustness, teel2003unified, yang2016stability, artstein2017asymptotic}. In these results, the derivative of the slow state is averaged over a finite or infinite time interval and the behavior of the reduced averaged slow system, together with the behavior of the boundary layer system, is used to describe the behavior of the full-order system. 
This idea can be found in the work of Gaitsgory \emph{et al.} \cite{gaitsgory1993suboptimal, gaitsgory2006averaging, gaitsgory2015averaging}, Grammel \cite{grammel1996singularly, grammel1997averaging, grammel2008robustness}, Artstein \emph{et al.} \cite{artstein1997tracking, artstein1998stability}, Teel \emph{et al.} \cite{teel2003unified}, and others \cite{wang2012analysis, yang2016stability}.

The problem of exponential stability of this general class of singular perturbation is not well studied in the literature. Among the above-mentioned results, Grammel showed in \cite{grammel2008robustness} that under the exponential stability of the origin of the reduced average system and under some other conditions on the system model, the slow state of a delayed singularly perturbed system is exponentially stable. However, the behavior of the fast state and also the closeness of solutions of the singularly perturbed system to the solutions of the reduced average and boundary layer systems when the reduced average system is not exponentially stable are not studied in \cite{grammel2008robustness}. 
%Furthermore, some of the assumptions in \cite{grammel2008robustness} are strong and not satisfied even by linear systems. 

This paper assumes a more general class of non-delayed singularly perturbed systems, compared to \cite{grammel2008robustness}, and presents closeness of solution results.
In particular, it is shown that under the exponential stability of the boundary layer system and some other conditions on the system model and over a finite time interval, the solutions to the singularly perturbed system are approximated by the solutions of the reduced average and boundary layer systems when the perturbation parameter, $\varepsilon$, is small. 
Although Grammel did not study closeness of solutions in \cite{grammel2008robustness}, Teel \emph{et. al} presented a closeness of solution result in \cite{teel2003unified} which can be applied to a more general class of singular perturbation systems. However, the order of magnitude of error is not studied in \cite{teel2003unified}. 
Compared to \cite{teel2003unified}, we propose stronger conditions on the system model and obtain stronger closeness of solution results; we show the approximation errors are of order $\mathcal{O}(\sqrt\varepsilon)$. 

%------------------------
\vspace{0.1cm}
\emph{Notation:} 
\begin{itemize}
\item $\|z\|_\eta$ denotes the distance between a point $z$ and a bounded set $\eta$ in $\mathbb{R}^m$, i.e.
\begin{align}
 \|z\|_\eta = \mbox{dist}(z,\eta)=\inf_{\mathpzc y\in\eta}\Vert  z-\mathpzc y\Vert.
\end{align}
\item A continuous function $\gamma$ : $R_{\geq0} \to R_{>0}$ is of class $\mathcal L$ (i.e. $\gamma\in\mathcal L$) if $\gamma(s)$ is positive and is strictly decreasing to zero as $s\to\infty$.
\item A continuous function $\alpha:\mathbb{R}_{\geq0}\to\mathbb{R}_{\geq0}$ is of class $\mathcal{K}_\infty$ if it is strictly increasing, $\alpha(0)=0$ and $\alpha(r)\to\infty$ as $r\to\infty$.
\item A function $\delta_1(\varepsilon)$ is of order $\mathcal{O}(\delta_2(\varepsilon))$, i.e. $\delta_1(\varepsilon)=\mathcal{O}(\delta_2(\varepsilon))$, if there exist positive constants $k$ and $c$ such that \cite[Definition 10.1]{khalil2002nonlinear}
\begin{align}\label{eq:def_oeps}
|\delta_1(\varepsilon)|\leq k|\delta_2(\varepsilon)|, \quad\forall |\varepsilon|<c.
\end{align} 
If $\delta_1(\varepsilon)$ and $\delta_2(\varepsilon)$ are continuous at $\varepsilon=0$, then \eqref{eq:def_oeps} implies that
\begin{align}\label{eq:cond_opes}
 \lim_{\varepsilon\to0}\frac{|\delta_1(\varepsilon)|}{|\delta_2(\varepsilon)|}\leq k <\infty.
\end{align}
\end{itemize}
%=====================================
\section{Preliminaries}\label{sec:preliminaries}
Consider a singularly perturbed system
\begin{subequations}
\label{eq:singular}
\begin{align}
 \dot x &= f(x,z,\varepsilon), \quad x(0)=x_0,\label{eq:singular1}\\
 \varepsilon\dot z &= g(x,z,\varepsilon), \quad z(0)=z_0, \label{eq:singular2}
\end{align}
\end{subequations}
where $\varepsilon>0$ is a small perturbation parameter, and $x\in\mathbb{R}^n$ and $z\in\mathbb{R}^m$ are respectively the slow and fast variables.
Define the \emph{fast-time} variable $\tau=t/\varepsilon$. Then in the $\tau$-domain, \eqref{eq:singular} can be written as
\begin{subequations}
\label{eq:fast}
\begin{align}
\frac{dx}{d\tau} &= \varepsilon f(x, z, \varepsilon),\label{eq:fast1}\\
\frac{dz}{d\tau} &= g(x, z, \varepsilon).\label{eq:fast2}
\end{align}
\end{subequations}
Letting $\varepsilon = 0$, \eqref{eq:fast1} becomes $dx/d\tau = 0$ which implies that the slow variable $x$ is fixed, i.e. $x(\tau)=x_0$, $\forall \tau\geq0$. Then the \emph{boundary-layer system} is obtained by setting $\varepsilon=0$ in \eqref{eq:fast2} as
\begin{equation}
\label{eq:boundary}
\frac{dz_b}{d\tau} = g(x_0, z_b, 0), \quad z_b(0)=z_0,
\end{equation}
where $z_b$ denotes the state of the boundary layer system, and $x_0$ is treated as a fixed parameter. 

Let $x_0\in B_R(0)$, $z_0\in M$, and $\varepsilon\in[0,\varepsilon_1]$ where  $B_R(0)\in\mathbb{R}^n$ denotes a ball of radius $R>0$ centered at the origin, $M$ denotes a compact set in $\mathbb{R}^m$ and $\varepsilon_1>0$. 
Unlike the classical singular perturbation problem, we assume the solutions to the boundary layer system, denoted by $\phi_b(\tau, x_0, z_0)$, $\forall x_0\in B_R(0)$, $z_0\in M$, or by $\phi_b(\tau)$ for the ease of notation, do not converge to a unique equilibrium, but converge to a bounded set. For example, the solutions to the boundary layer system may converge to a limit cycle.

We make the following assumptions.
%-------------
\vspace{0.1cm}
\begin{assumption}[Lipschitz continuity of $f$ and $g$]\label{ass:lipschitz}
The functions $f(x,z,\varepsilon)$ and $g(x,z,\varepsilon)$ are locally Lipschitz continuous in $(x,z,\varepsilon)$ $\in B_R(0)\times M \times [0,\varepsilon_1]$. We denote $L>0$ as the Lipschitz constant of $f(x,z,\varepsilon)$ and $g(x,z,\varepsilon)$ on $B_R(0)\times M\times  {[0,\varepsilon_1]}$.
\end{assumption}
%-------------
\vspace{0.1cm}
\begin{remark}[Bounds on $f$ and $g$]\label{rem:p}
From Assumption \ref{ass:lipschitz}, we obtain that for any compact set $B_R(0) \times M \times [0,\varepsilon_1]$, there exists an upper bound on $f(x,z,\varepsilon)$ and $g(x,z,\varepsilon)$; i.e. there exists $P>0$ such that 
\begin{align}\label{eq:p}
\|f(x,z,\varepsilon)\| \leq P, \quad \|g(x,z,\varepsilon)\| \leq P, \quad
\end{align}
 for all $(x,z,\varepsilon)\in B_R(0) \times M \times [0,\varepsilon_1]$.
\end{remark}
\vspace{0.1cm}
%------------- Assumption: forward invariance
\begin{assumption}[Forward invariance]\label{ass:invar}
There exists a positive constant $\varepsilon_1>0$ such that $B_R(0)\times M$ is forward invariant with respect to \eqref{eq:singular}
for all $\varepsilon\in[0,\varepsilon_1]$. 
Moreover, $B_R(0)$ is invariant with respect to
\begin{align}
 x(t) = \bar{x} + \int_0^{\bar{t}} f\big(\bar{x}, y(s),0\big)ds
\end{align} 
for all $\bar{t}>0$, where $y(s)$ is the solution to
\begin{align}
 \frac{dy(s)}{ds} = g\big(\bar{x}, y(s),0\big)
\end{align}
and $\bar{x}\in B_R(0)$ is a fixed parameter.
\end{assumption}
\vspace{0.1cm}
%------------------
In order to define the reduced average system, we will assume that $f\big(x,\phi_b(\tau),0\big)$ has a well-defined average. To be more precise, we state the following assumption that imposes conditions on $f$ such that the average of $f$ exists. The conditions in this assumption are similar to the conditions in \cite[Definition 10.2]{khalil2002nonlinear}.
%-------------
\vspace{0.1cm}
\begin{assumption}\label{ass:boundary}
The trajectories of the boundary layer system \eqref{eq:boundary} starting from $z_0\in M\subset \mathbb{R}^m$, denoted by $\phi_b(\tau,x,z_0),$ converge exponentially fast to a bounded set $\eta: \eta \in M$ which is possibly parametrized by $x$. The limit 
\begin{equation}\label{eq:fav}
 f_{av}(x) := \lim_{\mathpzc T\rightarrow\infty}\frac{1}{\mathpzc T}\int_{0}^{\mathpzc T} f\big(x,\phi_b(s,x,z_0),0\big)ds,
\end{equation}
exists and is the same for all $z_0\in M$. There exist $s^*>0$, $\gamma(s)\in\mathcal L$ and $\alpha(\cdot)\in\mathcal{K}_\infty$ such that
\begin{align}\label{eq:ineq}
&\frac{1}{s}\left\Vert\int_{\tau'}^{\tau'+s} \Big( f\big(x,\phi_b(\tau,x,z_0),0\big)-f_{av}(x) \Big) d\tau \right\Vert\nonumber\\
&\quad\leq \gamma(s)\alpha\left(\max\{\|x\|,\|z_0\|\}\right)
\end{align}
holds for all $\tau'\geq0$, $s> s^*$ and all boundary layer solutions 
$\phi_b(\tau,x,z_0)$ staring from an initial condition $z_0$ in $M$  for $\tau\in[\tau', \tau'+s]$. Here, $x$ is treated as a fixed parameter.
\end{assumption}
\vspace{0.1cm}
%------------ End assumption
Note that since $x$ and $z_0$ are assumed to be in compact sets $B_R(0)$ and $M$, the term $\alpha\left(\max\{\|x\|,\|z_0\|\}\right)$ on the right hand side of \eqref{eq:ineq} could be removed if we assume $\gamma(s)$ depends on $R$ and $M$. We used the above notation to emphasize the fact that the right hand side of \eqref{eq:ineq} is in general a function of $\|x\|$ and $\|z_0\|$.

If Assumption~\ref{ass:boundary} holds, we say $f\big(x,\phi_b(\tau),0\big)$ has a well-defined average $f_{av}(x)$. Then the \emph{reduced average system} (or what is called the \emph{reduced system} in the rest of the paper) is defined as
\begin{equation}
\label{eq:average}
\dot{x}_{av} = f_{av}(x_{av}), \quad x_{av}(0)=x_0.
\end{equation}

%------ Remark: Differential inclusion
\vspace{0.1cm}
\begin{remark}
In general, the reduced system should be defined as a differential inclusion of the form 
\begin{align}
 &\dot{x}_{av} \in F_{av}(x_{av}),\nonumber
\end{align}
where 
{\small
\begin{align}
 &F_{av}(x) = \mathrm{conv}\left( \bigcup_{z_0\in M}\left\{ \lim_{\mathpzc T\rightarrow\infty}\frac{1}{\mathpzc T}\int_{0}^{\mathpzc T} f\big(x,\phi_b(s,x,z_0),0\big)ds \right\}\right),\nonumber
\end{align}
}
with $\mathrm{conv}(S)$ denoting the closed convex hull of a set $S$.
This is due to the fact that $f_{av}$ in \eqref{eq:fav} is in general a function of $x$ and $z_0$; see e.g. \cite{grammel1997averaging, artstein1998stability}. 
We however assumed in this paper that the set valued map $F_{av}(x)$ is a singleton, i.e. $F_{av}(x) = \{f_{av}(x)\}$; see Assumption~\ref{ass:boundary}. 
This is a more restrictive assumption compared to \cite{grammel1997averaging, artstein1998stability} and more general conditions will be the topic for further research.
Therefore, we use the differential equation notation of \eqref{eq:average} for the reduced system.
\end{remark}
We finally make the following assumption on $f_{av}$.
\begin{assumption}\label{assump:Lipschitz_global}
The function $f_{av}(\cdot)$ is globally Lipschitz with Lipschitz constant $L_{av}>0$.
\end{assumption}
%=====================================
\section{Main result}\label{sec:main}
In this subsection, we analyze the closeness of solutions of the  singularly perturbed system and the reduced and boundary layer systems over a finite time interval. This result is  independent of any stability properties of the reduced system \eqref{eq:average}.

We aim to investigate the system on a finite time horizon $t\in[0,T]$ where $T>0$ and $t_0:=0$. We divide this time interval into sub intervals of the form $[t_l, t_{l+1}]$ which all have the same length $\varepsilon S_\varepsilon$, except possibly the last interval with length smaller than or equal to the length $\varepsilon S_\varepsilon$, and the index $l$ is an element of the index set $I_\varepsilon = \{0, 1, \cdots, \left\lfloor T/\varepsilon S_\varepsilon\right\rfloor\}$, where $\left\lfloor{\cdot}\right\rfloor$ denotes the floor function. The last time in the sequence is equal to $T$. In the following lemma, we define the mapping $S_\varepsilon$ and state some of its properties. The reason why this specific mapping is used will become clear later in the proof of  Lemma~\ref{lemma:DDelta} and Theorem~\ref{theorem:closeness}.

\vspace{0.1cm}
\begin{lemma}\label{lemma:se}
For any given $L>0$ and $T>0$, the map $\varepsilon\to S_\varepsilon$ defined as\footnote{This definition is inspired from \cite{grammel1997averaging}.}
\begin{align}\label{eq:se_definition}
 \frac{1}{{\varepsilon}^{1/4}}:=S_\varepsilon e^{T L\big(1+ S_\varepsilon L e^{LS_\varepsilon}\big)}
\end{align}
has the following properties 
\begin{subequations}\label{eq:se}
\begin{align}
 &\lim_{\varepsilon\to0} S_\varepsilon = \infty, \label{eq:se1}\\ 
 &\lim_{\varepsilon\to0} \varepsilon^{1/4} S_\varepsilon = 0.\label{eq:se2}
\end{align}
\end{subequations}
\end{lemma}
\vspace{0.1cm}
The proof of the above Lemma is given in the Appendix.

Denote the solution of \eqref{eq:singular} for $t\in[0,T]$ by $\big(x(t),z(t)\big)$ and define $\xi(t)$ for $t\in[t_l,t_{l+1}]$ as
\begin{align}\label{eq:xi}
 \xi(t):=\xi_l+\int_{t_l}^{t} f(\xi_l,y(s),0)ds,
\end{align}
with $\xi_l:=\xi(t_l)$ and $\xi_0=x(0)=x_0$, where $y(t):[t_l,t_{l+1}]\to \mathbb R^m$ is the unique solution to 
\begin{align}\label{eq:y}
 \varepsilon \dot{y}(t) = g(\xi_l,y(t),0), \quad y(t_l)=z(t_l).
\end{align}
Define
\begin{align}
 \Delta_l(t) &:= \max_{t_l\leq s \leq t}\|x(s)-\xi(s)\|,\label{eq:deltal_def}\\
 d_l(t) &:= \max_{t_l\leq s \leq t}\|x(s)-\xi_l\|,\label{eq:dsl_def}\\
 D_l(t) &:= \max_{t_l\leq s \leq t}\|z(s)-y(s)\|,\label{eq:dl_def}
\end{align}
for  $t\in[t_l, t_{l+1}]$. 
We state the following lemma for later use. The idea for the lemma is taken from \cite{grammel1997averaging}.
%----------------------------
\vspace{0.1cm}
\begin{lemma}\label{lemma:DDelta}
Consider the map $\varepsilon\to S_\varepsilon$ defined in Lemma~\ref{lemma:se} and suppose there exists a compact set $B_R(0)\times M \times (0,\varepsilon_1]$ on which Assumptions~\ref{ass:lipschitz} and \ref{ass:invar} hold. Then for any finite $T>0$ and for $t\in[0,T]$, the signals $ \Delta_l(t)$ and $D_l(t)$, $l\in I_\varepsilon$, defined respectively in \eqref{eq:deltal_def} and \eqref{eq:dl_def} are upper bounded by $\bar{\Delta}(\varepsilon)$ and $\bar{D}(\varepsilon)$ defined as
\begin{align}
 \bar{\Delta}(\varepsilon) &:= \Big( 2\varepsilon S_\varepsilon  P + T L(\varepsilon S_\varepsilon  P +\varepsilon)
 \left(1+LS_\varepsilon e^{LS_\varepsilon}\right)\Big)\label{eq:Deltabar}\nonumber\\
 &\qquad e^{T L\big(1+ S_\varepsilon L e^{LS_\varepsilon}\big)},\\
 \bar D(\varepsilon) &:= S_\varepsilon L \big(\bar{\Delta}(\varepsilon)+\varepsilon S_\varepsilon P+\varepsilon\big)e^{LS_\varepsilon}.\label{eq:barD}
\end{align}
Furthermore, $\bar{\Delta}(\varepsilon)$ and $\bar{D}(\varepsilon)$ are $\mathcal{O}({\sqrt{\varepsilon}})$.
\end{lemma}
\vspace{0.1cm}
%----------- Proof 
\begin{proof}
Refer to the Appendix for the proof.
\end{proof}
%------------------- Theorem: Closeness of solutions
\vspace{0.1cm}
\begin{theorem}[closeness of solutions over a finite time]\label{theorem:closeness}
Consider $S_\varepsilon$ defined in Lemma~\ref{lemma:se}. Suppose there exist $R>0$, $\varepsilon_1>0$ and a compact set $M$
such that Assumptions~\ref{ass:lipschitz}-\ref{assump:Lipschitz_global} hold on  $(x,z,\varepsilon)\in B_R(0)\times M \times (0,\varepsilon_1]$. Then for any finite time interval $t\in[0,T]$, 
\begin{enumerate}[(i)]
\item the solutions of the singularly perturbed system \eqref{eq:singular} and the reduced system \eqref{eq:average} satisfy
\begin{align}
&\|x(t)-x_{av}(t)\| \leq K(\varepsilon),
\end{align}
where $\lim_{\varepsilon\to0}K(\varepsilon)=0$ and $\|z(t)\|_\eta$ converges to an $F(\varepsilon)$ neighborhood of the bounded set $\eta$ exponentially fast where $F(\varepsilon):\mathbb{R}_{>0}\to\mathbb{R}_{>0}$ satisfies $\lim_{\varepsilon\to0}F(\varepsilon)=0$.

\item If we further assume there exist $\varepsilon^* \in (0,\varepsilon_1]$, $r'>0$ and  $\alpha_1: \alpha_1>2$ such that for $S_\varepsilon\geq S_{\varepsilon^*}$, the class-$\mathcal L$ function $\gamma(S_\varepsilon)$ satisfies
\begin{align}\label{eq:gamma_cond}
\gamma(S_\varepsilon) \leq r'e^{-\alpha_1 T L\big(1+ S_\varepsilon L e^{LS_\varepsilon}\big)},
\end{align}
then 
\begin{align}\label{eq:xOeps}
&\|x(t)-x_{av}(t)\| = \mathcal{O}(\sqrt\varepsilon)
\end{align}
holds for $\varepsilon\in(0,\varepsilon^*]$, uniformly on $t\in[0,T]$. Moreover, given any $t_a: 0<t_a<T$, there exists $\varepsilon^{**}\leq\varepsilon^*$ such that 
\begin{align}
 \Big|\|z(t)\|_\eta - \|\varphi_b(t/\varepsilon)\|_\eta\Big| = \mathcal{O}(\sqrt{\varepsilon})
\end{align}
holds uniformly on $t\in[t_a,T]$ when $\varepsilon\in(0,\varepsilon^{**}]$. 
\end{enumerate}
\end{theorem}
\vspace{0.1cm}
%-------------
\begin{proof} (i)
By Assumption~\ref{ass:boundary}, there exist positive constants $r_y$ and $\beta_y$ such that the solutions to the boundary layer system \eqref{eq:boundary} satisfy
\begin{align}\label{eq:sol_boundary}
 \|\phi_b(t/\varepsilon)\|_\eta \leq r_y e^{-\beta_y t/\varepsilon}\|z_0\|_\eta.
\end{align}
Define $\omega(t)$, $t\in[t_l, t_{l+1}]$, $l\in I_\varepsilon$  as
\begin{align}\label{eq:omega}
 \omega(t) &= \omega(t_l)+\int_{t_l}^t f_{av}(\xi_l)ds \nonumber\\
 &= \omega(t_l)+f_{av}(\xi_l)(t-t_l),
\end{align}
where $\omega(t_l)= x_{av}(t_l)$ and $\xi_l=\xi(t_l)$ is defined in \eqref{eq:xi}. We start with the slow state and estimate an upper bound for $\|x(t)-x_{av}(t)\|$,
\begin{align}
 \|x(t)-x_{av}(t)\| &\leq \|x(t)-\xi(t)\|+\|\xi(t)-\omega(t)\|\nonumber\\
 &\quad+\|\omega(t)-x_{av}(t)\|.\label{eq:xnormbc}
\end{align}
From \eqref{eq:deltal_def} and Lemma~\ref{lemma:DDelta}, for any $l$ in the index set $I_\varepsilon$, the first term on the right hand side of \eqref{eq:xnormbc} is less than or equal to $\bar{\Delta}(\varepsilon)$. Using Assumption~\ref{ass:boundary} and the fact that $\xi(0)=\omega(0)=x_0$, the second term can be written as
\begin{align}
 \|\xi(t)&-\omega(t)\| \leq  \|\xi(t_l)-\omega(t_l)\| \nonumber\\ 
 & \hspace{0.3cm}+ \left\Vert \int_{t_l}^{t}\big(f(\xi_l,y(s),0)-f_{av}(\xi_l)\big)ds\right\Vert \nonumber \\
 &\leq \|\xi(t_l)-\omega(t_l)\| + \varepsilon S_\varepsilon\gamma(S_\varepsilon)\max\{\|\xi(t_l)\|,\|z(t_l)\|\}\nonumber\\
 &\leq T\gamma(S_\varepsilon)\max\{\|\xi(t_l)\|,\|z(t_l)\|\}\nonumber\\
 &\leq T\gamma(S_\varepsilon)\max\{R,\bar{z}\},
\end{align}
%where $\bar{\gamma}(S_\varepsilon)$ is a class $\mathcal{L}$ function, and 
where $\bar{z} = \max_{ z\in M}{\|z\|}$. 
%From the definition of $f_{av}(\cdot)$ in \eqref{eq:fav} and the boundedness of $f$ on $B_R(0) \times M \times [0,\varepsilon_1]$, the third term can be upper bounded by
%\begin{align}
% \|\omega&(t)-x_{av}(t)\| \nonumber\\ 
% & \leq  \|\omega(t_l)-x_{av}(t_l)\| + \left\Vert \int_{t_l}^{t}\Big(f_{av}(\xi_l)-f_{av}\big(x_{av}(s)\big)\Big)ds \right\Vert\nonumber\\
% &= \left\Vert \int_{t_l}^{t}\Big(f_{av}(\xi_l)-f_{av}\big(x_{av}(s)\big)\Big)ds \right\Vert \leq  2\varepsilon S_\varepsilon P.
%\end{align}
Using Assumption~\ref{assump:Lipschitz_global} and the Gronwall-Bellman inequality \cite[Lemma~A.1]{khalil2002nonlinear}, the third term can be upper bounded by
\begin{align}
 \|\omega&(t)-x_{av}(t)\| \nonumber\\ 
 & \leq  \|\omega(t_l)-x_{av}(t_l)\| + \left\Vert \int_{t_l}^{t}\Big(f_{av}(\xi_l)-f_{av}\big(x_{av}(s)\big)\Big)ds \right\Vert\nonumber\\
 &= \left\Vert \int_{t_l}^{t}\Big(f_{av}(\xi_l)-f_{av}\big(x_{av}(s)\big)\Big)ds \right\Vert \nonumber\\
 &\leq L_{av} \int_{t_l}^t \| \xi_l-x_{av}(s)\|ds \nonumber\\
 &\leq L_{av} \int_{t_l}^t \Big( \|\xi_l-\xi(s)\| + \|\xi(s)-\omega(s)\|\nonumber\\
 & \qquad + \|\omega(s)- x_{av}(s)\|\Big)ds \nonumber\\ 
 & \leq \varepsilon S_\varepsilon L_{av} \Big( \varepsilon S_\varepsilon P + T\gamma(S_\varepsilon)\max\{R,\bar{z}\}\Big)e^{\varepsilon S_\varepsilon L_{av}}.
\end{align}
Define $K(\varepsilon)$ as
\begin{align}\label{eq:kepsbc}
 K(\varepsilon)&:= \bar{\Delta}(\varepsilon)+T\gamma(S_\varepsilon)\max\{R,\bar{z}\}\\
 &\quad + \varepsilon S_\varepsilon L_{av} \Big( \varepsilon S_\varepsilon P + T\gamma(S_\varepsilon)\max\{R,\bar{z}\}\Big)e^{\varepsilon S_\varepsilon L_{av}}.
\end{align}
Then we obtain from \eqref{eq:xnormbc} and \eqref{eq:kepsbc} that for any finite time interval $[0,T]$,
\begin{align}
\|x(t)-x_{av}(t)\| &\leq K(\varepsilon).
\end{align}
Note that $K(\varepsilon)$ is uniform in $(x_0,z_0)\in B_R(0)\times M$, and from Lemma~\ref{lemma:se} and Lemma~\ref{lemma:DDelta}, $\lim_{\varepsilon\to0}K(\varepsilon)=0$.

%-------------------- Fast state
We now study the behavior of the fast state, $z(t)$. 
Using the triangle inequality, we obtain for $t\in[t_l, t_{l+1}]$ that
\begin{align}
\label{eq:z_boundbc}
 \|z(t)\|_\eta &\leq \|y(t)\|_\eta + \|z(t)-y(t)\|.
\end{align}
Note that $y(t)$ is the solution to \eqref{eq:y} and is different from $\phi_b(t/\varepsilon)$, the solution to the boundary layer system \eqref{eq:boundary}. Indeed, the signal $y(t)$ is defined such that its value at the time instant $t_l$, $l\in I_\varepsilon$ is equal to $z(t_l)$ and changes according to \eqref{eq:y} over the interval $[t_l, t_{l+1}]$. 
However, 
%we know from Assumption~\ref{ass:boundary} that the exponential stability of the boundary layer system holds uniformly in the frozen parameter, and therefore, 
the boundary-layer system \eqref{eq:y} can be represented as a boundary layer model of the form \eqref{eq:boundary} with $\xi_l$ as the frozen parameter. Hence the solution of \eqref{eq:y} for $t\in[t_l, t_{l+1}]$ satisfies the same inequality as \eqref{eq:sol_boundary}, with a different initial condition, for all $x$ and $\xi_l$ in $B_R(0)$. So we obtain from \eqref{eq:z_boundbc} and Lemma~\ref{lemma:DDelta} that 
\begin{align}
\label{eq:z_boundbc2}
 \|z(t)\|_\eta &\leq r_y e^{-\beta_y t/\varepsilon}\|y(t_l)\|_\eta + \bar{D}(\varepsilon)\nonumber\\
& \stackrel{\eqref{eq:y}}{=} r_y e^{-\beta_y t/\varepsilon}\|z(t_l)\|_\eta + \bar{D}(\varepsilon).
\end{align}
Specifically, we obtain for $t=t_{l+1}$ that 
\begin{align}
 \|z(t_{l+1})\|_\eta &\leq r_y e^{-\beta_y S_\varepsilon}\|z(t_l)\|_\eta + \bar{D}(\varepsilon).
\end{align}
Choose $\delta_y\in(0,\beta_y)$ and $\bar{\varepsilon}>0$ such that 
\begin{align}\label{eq:rz}
 e^{-\delta_y S_{\bar{\varepsilon}}}\leq \frac{1}{r_y}.
\end{align}
Then we obtain by inclusion for all $l\in I_\varepsilon$ and all $\varepsilon\in(0,\min\{\varepsilon_1,\bar{\varepsilon}\}]$ that 
\begin{align}\label{eq:zb_a}
 \|z(t_{l+1})\|_\eta &\leq e^{-(l+1)(\beta_y-\delta_y) S_\varepsilon}\|z_0\|_\eta + \bar{D}(\varepsilon)\sum_{k=0}^{l} e^{-k(\beta_y-\delta_y) S_\varepsilon}\nonumber\\
 &= e^{-(l+1)(\beta_y-\delta_y) S_\varepsilon}\|z_0\|_\eta \nonumber\\
 &\quad + \bar{D}(\varepsilon)\frac{1-e^{-(\beta_y-\delta_y)(l+1)S_\varepsilon}}{1-e^{-(\beta_y-\delta_y)S_\varepsilon}},
\end{align}
and obtain for $t\in[t_l,t_{l+1}]$ that
\begin{align}
 \|z(t)\|_\eta &\leq r_y e^{-\beta_y S_\varepsilon}\|z(t_l)\|_\eta + \bar{D}(\varepsilon)\nonumber\\
\To{\eqref{eq:zb_a}}\quad & \leq r_y e^{-\beta_y S_\varepsilon} e^{-l(\beta_y-\delta_y) S_\varepsilon}\|z_0\|_\eta \nonumber\\
 &\quad + \bar{D}(\varepsilon) r_y e^{-\beta_y S_\varepsilon}\frac{1-e^{-(\beta_y-\delta_y)lS_\varepsilon}}{1-e^{-(\beta_y-\delta_y)S_\varepsilon}} + \bar{D}(\varepsilon)\nonumber\\
  & \leq r_y e^{-(\beta_y-\delta_y) t/\varepsilon}\|z_0\|_\eta \nonumber\\
 &\quad + \bar{D}(\varepsilon) r_y e^{-\beta_y S_\varepsilon}\frac{1-e^{-(\beta_y-\delta_y)lS_\varepsilon}}{1-e^{-(\beta_y-\delta_y)S_\varepsilon}} + \bar{D}(\varepsilon),\label{eq:zb_bc}
\end{align}
where we used $l=t_l/(\varepsilon S_\varepsilon)$ and $t_l\leq t\leq t_{l+1}$.
Define $F(\varepsilon)$ as 
\begin{align}\label{eq:Fepsc}
F(\varepsilon):= \bar{D}(\varepsilon)\left(1 +  \frac{r_y e^{-\beta_y S_\varepsilon}}{1-e^{-(\beta_y-\delta_y)S_\varepsilon}} \right).
\end{align}
Then we obtain that 
\begin{align}
\|z(t)\|_\eta &\leq r_y e^{-(\beta_y-\delta_y)t/\varepsilon}\|z_0\|_\eta + F(\varepsilon).\label{eq:pr3bc}
\end{align}
where $\lim_{\varepsilon\to0} F(\varepsilon)=0$. 
The proof of the first part of the theorem is complete.

(ii)
In the second part of the proof, we first show that under \eqref{eq:gamma_cond}, $K(\varepsilon)=\mathcal{O}(\sqrt\varepsilon)$. 
From Lemma~\ref{lemma:DDelta}, $\bar{\Delta}(\varepsilon)$ which is the first term on the right hand side of \eqref{eq:kepsbc} is of order $\mathcal{O}(\sqrt\varepsilon)$. For the the second term we have 
\begin{align}
&\lim_{\varepsilon\to0}\frac{T\gamma(S_\varepsilon)\max\{R,\bar{z}\}}{\sqrt{\varepsilon}} \nonumber\\
&\stackrel{\eqref{eq:se_definition}}{=} \lim_{\varepsilon\to0} T\max\{R,\bar{z}\} \gamma(S_\varepsilon)S_\varepsilon^2 e^{2T L\big(1+ S_\varepsilon L e^{LS_\varepsilon}\big)}\nonumber\\
&\To{\eqref{eq:gamma_cond}}\quad\leq  T\max\{R,\bar{z}\} \lim_{\varepsilon\to0} S_\varepsilon^2 e^{-(\alpha_1-2)T L\big(1+ S_\varepsilon L e^{LS_\varepsilon}\big)}\nonumber\\
&\quad\quad~~=0. \label{eq:gamalimt}
\end{align}
The last term is also of order $\mathcal{O}(\sqrt{\varepsilon})$. 
So \eqref{eq:xOeps} holds uniformly for $t\in[0,T]$ when $0<\varepsilon\leq\varepsilon^*$ where $\varepsilon^*$ satisfies 
\begin{align}
 \frac{1}{{\varepsilon^*}^{1/4}}:=S_{\varepsilon^*} e^{T L\big(1+ S_{\varepsilon^*} L e^{LS_{\varepsilon^*}}\big)}.
\end{align}
From \eqref{eq:Fepsc} and the fact that $\bar{D}(\varepsilon)=\mathcal{O}(\sqrt\varepsilon)$, see Lemma~\ref{lemma:DDelta}, $F(\varepsilon)$ is also of order $\mathcal{O}(\sqrt\varepsilon)$ as 
\begin{align}
\lim_{\varepsilon\to0}\left(1 +  \frac{r_y e^{-\beta_y S_\varepsilon}}{1-e^{-(\beta_y-\delta_y)S_\varepsilon}} \right) = 1<\infty.
\end{align}
From \eqref{eq:sol_boundary} and \eqref{eq:pr3bc}, we have
\begin{align}
 &\Big|\|z(t)\|_\eta - \|\varphi_b(t/\varepsilon)\|_\eta\Big|\nonumber\\
 &\qquad\leq r_y e^{-(\beta_y-\delta_y)t/\varepsilon}\|z_0\|_\eta + F(\varepsilon)\nonumber + r_y e^{-\beta_yt/\varepsilon}\|z_0\|_\eta\nonumber\\
%-----------
 &\qquad\leq 2r_y e^{-(\beta_y-\delta_y)t/\varepsilon}\|z_0\|_\eta+F(\varepsilon).
\end{align}
Then since 
\begin{align}
e^{-(\beta_y-\delta_y)t/\varepsilon}\leq \sqrt{\varepsilon}, 
\quad \forall (\beta_y-\delta_y)t \geq \varepsilon\ln(\frac{1}{\sqrt{\varepsilon}}),
\end{align}
we can choose $\varepsilon^{**}$ such that 
\begin{align}
(\beta_y-\delta_y)t_a = \varepsilon^{**}\ln(\frac{1}{\sqrt{\varepsilon^{**}}}),
\end{align}
and we conclude that 
\begin{align}
\Big|\|z(t)\|_\eta - \|\varphi_b(t/\varepsilon)\|_\eta\Big| = \mathcal{O}(\sqrt\varepsilon)
\end{align}
holds uniformly on $t\in[t_a,T]$ for $0<\varepsilon\leq\varepsilon^{**}$.
\end{proof}
\section{Simulations}\label{sec:simulations}
\begin{figure}[h]
  \center
  % left bottom right top
  \includegraphics[trim = 0mm 0mm 0mm 0mm, width=0.45\textwidth]{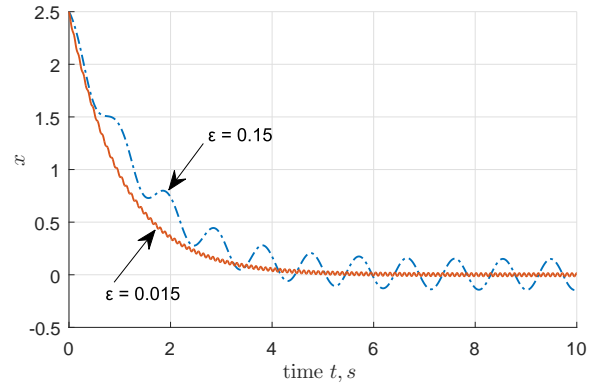}
  \caption{The slow variable $x(t)$ of the full-order system \eqref{eq:singular} for different values of $\varepsilon$.}
  \label{fig:sim_x}
\end{figure}

\begin{figure}[h]
  \centering
  \includegraphics[width=.45\textwidth]{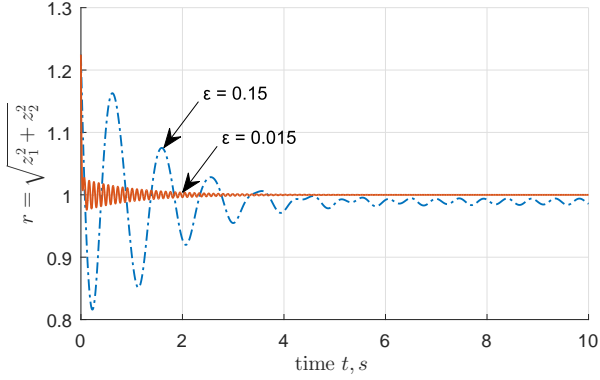}
\caption{The norm of the fast variable of the full-order system, $\|z(t)\|$.}  \label{fig:sim_z}
\end{figure}

%---------- Example 1
In this section, we present a numerical example in which the solution of the boundary layer system converges to a limit cycle. Consider the following system 
\begin{align}
 \dot{x} &= -x + z_1 + \varepsilon x^2 \nonumber\\
 \varepsilon\dot{z_1} &= -z_1+z_2+\frac{z_1}{\sqrt{z_1^2+z_2^2}} \label{ex:sys}\\
 \varepsilon\dot{z_2} &= -z_1-z_2+\frac{z_2}{\sqrt{z_1^2+z_2^2}}+\varepsilon x, \nonumber
\end{align}
and note the system is not defined for $z_1=z_2=0$ and thus the subset $M$ does not include the origin. Defining $r$ and $\theta$ such that $z_1=r\cos\theta$ and $z_2=r\sin\theta$, the state equations \eqref{ex:sys} can be written in the polar coordinates as 
\begin{align}
 \dot{x} &= -x + r\cos\theta +\varepsilon x^2 \nonumber\\
 \varepsilon\dot{r} &= 1-r+\varepsilon x\sin\theta\label{ex:sys_pol}\\
 \varepsilon\dot{\theta} &= -1+\varepsilon x\frac{\cos\theta}{r}.\nonumber
\end{align}
Define $z=[z_1~z_2]^\top$ and define the isolated periodic orbit $\eta$ as 
$$\eta = \{z\in \mathbb R^2 \mid \|z\|=1 \}.$$
Then 
$$\|z\|_\eta = \mbox{dist}(z,\eta)=\inf_{y\in\eta} \|z-y\| = \left\vert \|z\|-1 \right\vert.$$
Letting $\varepsilon=0$ in \eqref{ex:sys}, the boundary layer system can be written as
\begin{align}
 \frac{dz_1}{d\tau} &= -z_1+z_2+\frac{z_1}{\sqrt{z_1^2+z_2^2}} \nonumber\\
 \frac{dz_2}{d\tau} &= -z_1-z_2+\frac{z_2}{\sqrt{z_1^2+z_2^2}}\nonumber
\end{align}
which is equivalent (for $r>0$) to 
\begin{align}
 \frac{dr}{d\tau} &= 1-r, \qquad
 \frac{d\theta}{d\tau} = -1\nonumber
\end{align}
in polar coordinates. Thus for $r>0$, the orbit $r=\|z\|=1$ is exponentially stable and the solution to the boundary layer system is
\begin{align}
 z_1(\tau) &= \big( (r_0-1)e^{-\tau}+1\big)\cos(-\tau+\theta_0) \nonumber\\
 z_2(\tau) &= \big( (r_0-1)e^{-\tau}+1\big)\sin(-\tau+\theta_0)\nonumber
\end{align}
where $\theta_0=\mathrm{atan}(z_2(0)/z_1(0))$ and $r_0=\|z_0\|=\sqrt{z_1^2(0)+z_2^2(0)}$. This solution can also be written as
$$\|z(\tau)\|_\eta = e^{-\tau}\|z_0\|_\eta.$$
From \eqref{eq:fav} and \eqref{eq:average}, the reduced system is defined as
\begin{align}
 \dot{x}_{av} &= f_{av}(x_{av}) = -x_{av}  \nonumber\\
 &+\lim_{\mathpzc T\rightarrow\infty}\frac{1}{\mathpzc T}\int_{0}^{\mathpzc T} \big( (r_0-1)e^{-s}+1\big)\cos\big(-s+\theta_0\big)ds\nonumber\\
 &=-x_{av}.\nonumber
\end{align}
%Hence the reduced system is globally exponentially stable. 
We now check the validity of Assumption~\ref{ass:boundary}.
\begin{align}\label{eq:ineq2}
&\frac{1}{s}\left\Vert\int_{\tau'}^{\tau'+s} \Big( f\big(x,\phi_b(\tau),0\big)-f_{av}(x) \Big) d\tau \right\Vert \nonumber\\
%--------
&= \frac{1}{s}\left\vert\int_{\tau'}^{\tau'+s}  \big( (r_0-1)e^{-\tau}+1\big)\cos\big(-\tau+\theta_0\big)d\tau\right\vert\nonumber\\
%--------
&\leq \left\vert\frac{1}{s}\frac{r_0-1}{2}e^{-\tau}\big(-\sin(-\tau+\theta_0)-\cos(-\tau+\theta_0)\big)\right.\nonumber\\
&\quad-\left.\left.\frac{1}{s}\sin(-\tau+\theta_0)\right|^{\tau=\tau'+s}_{\tau=\tau'}\right|\\
%--------
&\leq \frac{2}{s} \max\{r_0,1\}.
\end{align}
Thus $\gamma(s)=2/s$ and Assumption~\ref{ass:boundary} holds for all $s^*>0$. Assumption~\ref{assump:Lipschitz_global} also holds.
Choose $\varepsilon_1=0.15$, $R=2.5$, and $M=\{z\in \mathbb{R}^2\backslash\{0\}\mid 0.5\leq\|z\|\leq 1.5\}$, and observe that Assumptions~\ref{ass:lipschitz} and \ref{ass:invar} hold on $(x,z,\varepsilon)\in B_R(0)\times M\times [0,\varepsilon_1)$.
So all conditions of Theorem~\ref{theorem:closeness} hold and therefore the solutions of the singularly perturbed system are approximated, for sufficiently small $\varepsilon>0$, by the solutions of the reduced average and boundary layer systems. This is shown in Fig.~\ref{fig:sim_x} and Fig.~\ref{fig:sim_z} where the trajectories of \eqref{ex:sys} are depicted for $\varepsilon = 0.15$ and $\varepsilon=0.015$.
%====================================================

%=============================== Conclusions
\section{Conclusion}\label{sec:conclusion}
In this paper, we have studied the behavior of a general singularly perturbed system with solutions of the boundary layer system converging exponentially fast to a bounded set. We used averaging to eliminate the fast oscillations of the fast state, and presented results on the closeness of solutions of the full-order system and the reduced average system over a finite time interval.
%================================ Appendix
\section{Appendix}
\subsection*{Proof of Lemma~\ref{lemma:se}.}
Consider the definition of $S_\varepsilon$ in \eqref{eq:se_definition}, and note that as $\varepsilon$ goes to zero, $S_\varepsilon e^{T L\big(1+ S_\varepsilon L e^{LS_\varepsilon}\big)}$ goes to infinity which implies that $S_\varepsilon$ goes to infinity. Therefore $\lim_{\varepsilon\to0}S_\varepsilon=\infty$. 

To show that $\lim_{\varepsilon\to0} \varepsilon^{1/4} S_\varepsilon=0$, observe that 
\begin{align}
 &\lim_{\varepsilon\to0}\varepsilon^{1/4} S_\varepsilon = \lim_{\varepsilon\to0} e^{-T L\big(1+ S_\varepsilon L e^{LS_\varepsilon}\big)}.
\end{align}
Then from $\lim_{\varepsilon\to0}S_\varepsilon=\infty$, we obtain that $\lim_{\varepsilon\to0}\varepsilon^{1/4} S_\varepsilon=0$.
\hfill$\Box$

%-------------------------------------------
\vspace{0.1cm}
\subsection*{Proof of Lemma~\ref{lemma:DDelta}.}
Consider $\Delta_l(t)$ and $d_l(t)$ defined in \eqref{eq:deltal_def} and \eqref{eq:dsl_def} and note there is a bound $P$ on the norm of $f$ according to Remark~\ref{rem:p}. Then for $t\in[t_l,t_{l+1}]$, we have 
\begin{align}
 d_l(t) &= \max_{t_l\leq s \leq t}\|x(s)-\xi_l\| = \max_{t_l\leq s \leq t}\|x(s)-\xi(s)+\xi(s)-\xi_l\| \nonumber\\
 &\leq \max_{t_l\leq s \leq t}\|x(s)-\xi(s)\| + \max_{t_l\leq s \leq t}\|\xi(s)-\xi_l\|\nonumber\\
 &\leq \Delta_l(t) + \max_{t_l\leq s \leq t} \int_{t_l}^s \|f(\xi_l,y(s),0)\|ds\nonumber\\
 &\leq \Delta_l(t) + \varepsilon S_\varepsilon P.\label{eq:delta_l}
\end{align}
From \eqref{eq:singular} and \eqref{eq:y} we have 
\begin{align}
 \|z(t)-y(t)\|&=\frac{1}{\varepsilon}\left\Vert \int_{t_l}^t \Big( g(x(s),z(s),\varepsilon)-g(\xi_l,y(s),0)\Big)ds\right\Vert\nonumber.
\end{align}
Then using the Lipschitz property of $g$ in Assumption~\ref{ass:lipschitz} and the Gronwall-Bellman inequality \cite[Lemma~A.1]{khalil2002nonlinear}, we obtain
\begin{align}
D_l(t) &= \max_{t_l\leq s \leq t}\frac{1}{\varepsilon}\left\Vert \int_{t_l}^s \Big( g(x(s),z(s),\varepsilon)-g(\xi_l,y(s),0)\Big)ds\right\Vert \nonumber\\
 &\leq \max_{t_l\leq s \leq t}\frac{L}{\varepsilon} \int_{t_l}^s \Big( \|x(s)-\xi_l\| + \|z(s)-y(s)\| +\varepsilon \Big)ds \nonumber\\
 \To{\eqref{eq:dsl_def}, \eqref{eq:dl_def}} &\leq S_\varepsilon L\big(d_l(t)+\varepsilon\big) + \frac{L}{\varepsilon} \int_{t_l}^t D_l(s) ds \nonumber\\
 &\leq S_\varepsilon L\big(d_l(t)+\varepsilon\big)e^{
LS_\varepsilon}.\label{eq:Delta_l}
\end{align}
From \eqref{eq:singular} and \eqref{eq:xi} we have 
\begin{align}
\max_{t_l\leq s\leq t}\|x(s)-\xi(s)\|&\leq\left\Vert x(t_l)-\xi_l\right\Vert\nonumber\\ 
& \hspace{-2.5cm}+ \max_{t_l\leq s \leq t}\left\Vert\int_{t_l}^s \big( f(x(s),z(s),\varepsilon)-f(\xi_l,y(s),0)\big)ds\right\Vert
\end{align}
and thus we obtain using \eqref{eq:delta_l}, \eqref{eq:Delta_l} and the Gronwall-Bellman inequality that
\begin{align}
\Delta_l(t) &\leq \Delta_l(t_l) + L \int_{t_l}^t \big( d_l(s)+D_l(s)+\varepsilon\big)ds \nonumber\\
   \To{\eqref{eq:Delta_l}} \quad &\leq \Delta_l(t_l) + L \int_{t_l}^t \big(d_l(s)+\varepsilon\big)\left(1+LS_\varepsilon e^{LS_\varepsilon}\right)ds \nonumber\\
  \To{\eqref{eq:delta_l}} \quad &\leq \Delta_l(t_l) + \varepsilon S_\varepsilon L\big(\varepsilon S_\varepsilon P+\varepsilon\big)\big(1+LS_\varepsilon e^{LS_\varepsilon}\big)\nonumber\\ 
 &\hspace{0.5cm} + L\big(1+LS_\varepsilon e^{LS_\varepsilon}\big) \int_{t_l}^t \Delta_l(s)ds \nonumber\\ 
 &\leq \Big(\Delta_l(t_l) + \varepsilon S_\varepsilon L\big(\varepsilon S_\varepsilon P+\varepsilon\big)\nonumber\\ &\hspace{0.5cm} \big(1+LS_\varepsilon e^{LS_\varepsilon}\big)\Big)
  e^{\varepsilon S_\varepsilon L\left(1+LS_\varepsilon e^{LS_\varepsilon}\right)}.
\end{align}
Specifically, for $t=t_{l+1}$ we have
\begin{align}
\Delta_l(t_{l+1}) &\leq \Big( \Delta_l(t_l) + \varepsilon S_\varepsilon L(\varepsilon S_\varepsilon P+\varepsilon)\nonumber\\
 & \hspace{0.6cm} \left(1+LS_\varepsilon e^{LS_\varepsilon}\right)\Big) e^{\varepsilon S_\varepsilon L\big(1+ S_\varepsilon L e^{LS_\varepsilon}\big)}.
\end{align}
From the definition of $\Delta_l(t_l)$ in \eqref{eq:deltal_def}, we have $\Delta_l(t_l)\leq \Delta_{l-1}(t_l)$ and 
\begin{align*}
\Delta_0(t_1) &= \max_{t_0 \leq s \leq t_1}\|x(s)-\xi(s)\| \nonumber\\
 &= \max_{t_0\leq s\leq t_1}\left\Vert \int_{0}^{s} \big( f(x(s),z(s),\varepsilon)-f(\xi_l,y(s),0)\big)ds \right\Vert \nonumber\\
 &\leq 2\varepsilon S_\varepsilon P,
\end{align*}
where we assumed a bound $P$ for the norm of $f$ according to Remark~\ref{rem:p}. Hence we conclude for all $l$ in $I_\varepsilon$ that
\begin{align}\label{eq:Delta}
\Delta_l(t) \leq \Delta_l(t_{l+1}) &\leq \bar{\Delta}(\varepsilon),
\end{align}
where $\bar{\Delta}(\varepsilon)$ is defined as \eqref{eq:Deltabar}.
%\begin{align}\label{eq:Deltabar}
% \bar{\Delta}(\varepsilon) &:= \Big( 2\varepsilon S_\varepsilon  P + T L(\varepsilon S_\varepsilon  P +\varepsilon)
% \left(1+LS_\varepsilon e^{LS_\varepsilon}\right)\Big)\nonumber\\
% &\quad e^{T L\big(1+ S_\varepsilon L e^{LS_\varepsilon}\big)}.
%\end{align}
Given \eqref{eq:delta_l}, \eqref{eq:Delta_l} and \eqref{eq:Delta}, we also obtain that
\begin{align}
  D_l(t)\leq D_l(t_{l+1})\leq \bar{D}(\varepsilon),
\end{align}
with $\bar{D}(\varepsilon)$ defined in \eqref{eq:barD}.
%\hfill$\Box$

To show that $\bar{\Delta}(\varepsilon)=\mathcal{O(\sqrt{\varepsilon})}$, we split the right hand side of \eqref{eq:Deltabar} into the the following three terms and show that they are all $\mathcal{O(\sqrt{\varepsilon})}$. We use \eqref{eq:cond_opes} to check the order of magnitude of each of these terms.
\begin{align}
 %----- 1
(i):\quad& \lim_{\varepsilon\to0} \frac{2\varepsilon S_\varepsilon P e^{T L\big(1+ S_\varepsilon L e^{LS_\varepsilon}\big)}}{\sqrt{\varepsilon}}\nonumber\\ 
 &\quad \stackrel{\eqref{eq:se_definition}}{=} \lim_{\varepsilon\to0} 2\varepsilon^{1/4} P=0,\label{eq:lim1}\\
 %----- 2
(ii):\quad& \lim_{\varepsilon\to0} \frac{1}{\sqrt{\varepsilon}} T L\varepsilon S_\varepsilon P \left(1+LS_\varepsilon e^{LS_\varepsilon}\right) e^{T L\big(1+ S_\varepsilon L e^{LS_\varepsilon}\big)}\nonumber\\ 
 &\quad \stackrel{\eqref{eq:se_definition}}{=} P \lim_{\varepsilon\to0} \frac{1}{\sqrt{\varepsilon}} \varepsilon S_\varepsilon \ln\left(\frac{1}{\varepsilon^{1/4}S_\varepsilon}\right) \frac{1}{\varepsilon^{1/4}S_\varepsilon}\nonumber\\
 &\quad = P \lim_{\varepsilon\to0} \frac{1}{S_\varepsilon}\varepsilon^{1/4} S_\varepsilon \ln\left(\frac{1}{\varepsilon^{1/4}S_\varepsilon}\right)\stackrel{\eqref{eq:se}}{=}0.\label{eq:lim2}
\end{align}
Here, we used the fact that $\lim_{x\to0} x\ln \frac{1}{x}=0$.
\begin{align}
 %----- 3
(iii):\quad &\lim_{\varepsilon\to0} \frac{1}{\sqrt{\varepsilon}} T L\varepsilon \left(1+LS_\varepsilon e^{LS_\varepsilon}\right) e^{T L\big(1+ S_\varepsilon L e^{LS_\varepsilon}\big)}\nonumber\\ 
 &\quad \stackrel{\eqref{eq:se_definition}}{=} \lim_{\varepsilon\to0}\frac{1}{\sqrt{\varepsilon}}\varepsilon \ln\left(\frac{1}{\varepsilon^{1/4}S_\varepsilon}\right) \frac{1}{\varepsilon^{1/4}S_\varepsilon}\nonumber\\
 &\quad = \lim_{\varepsilon\to0} \frac{1}{(S_\varepsilon)^2}\varepsilon^{1/4} S_\varepsilon \ln\left(\frac{1}{\varepsilon^{1/4}S_\varepsilon}\right)\stackrel{\eqref{eq:se}}{=}0.\label{eq:lim3}
\end{align}
We now show that $\bar{D}(\varepsilon)=\mathcal{O}(\sqrt\varepsilon)$. 
We obtain from \eqref{eq:se_definition} that
\begin{equation}\label{eq:se_2}
 S_\varepsilon Le^{LS_\varepsilon}=\frac{1}{TL}\ln\left(\frac{1}{\varepsilon^{1/4}S_\varepsilon}\right)-1.
\end{equation}
Similarly to the above calculations for $\bar{\Delta}(\varepsilon)$, it can be shown using \eqref{eq:se_2} that $(\varepsilon S_\varepsilon P+\varepsilon)S_\varepsilon Le^{LS_\varepsilon}=\mathcal{O}(\sqrt{\varepsilon})$. We show below that  $S_\varepsilon Le^{LS_\varepsilon}\bar{\Delta}(\varepsilon)=\mathcal{O}(\sqrt{\varepsilon})$. 
Equation \eqref{eq:se_2} implies that $S_\varepsilon Le^{LS_\varepsilon}\bar{\Delta}(\varepsilon)=\frac{1}{TL}\ln\left(\frac{1}{\varepsilon^{1/4}S_\varepsilon}\right)\bar{\Delta}(\varepsilon)-\bar{\Delta}(\varepsilon)$.
Given \eqref{eq:Deltabar}, we split $\ln\left(\frac{1}{\varepsilon^{1/4}S_\varepsilon}\right)\bar{\Delta}(\varepsilon)$ into the following three terms and show they are $\mathcal{O}(\sqrt\varepsilon)$
\begin{align}
 %----- 1
(i):\quad& \lim_{\varepsilon\to0} \frac{1}{\sqrt{\varepsilon}} 2\ln\left(\frac{1}{\varepsilon^{1/4}S_\varepsilon}\right)\varepsilon S_\varepsilon P e^{T L\big(1+ S_\varepsilon L e^{LS_\varepsilon}\big)}\nonumber\\ 
&\quad \stackrel{\eqref{eq:se_definition}}{=}  2P\lim_{\varepsilon\to0} \varepsilon^{1/4} \ln\left(\frac{1}{\varepsilon^{1/4}S_\varepsilon}\right)\nonumber\\
&\quad= 2P\lim_{\varepsilon\to0} \frac{1}{S_\varepsilon}\varepsilon^{1/4}S_\varepsilon \ln\left(\frac{1}{\varepsilon^{1/4}S_\varepsilon}\right)\stackrel{\eqref{eq:se}}{=}0,\label{eq:limb1}\\
 %----- 2
(ii):\quad& \lim_{\varepsilon\to0}\frac{1}{\sqrt{\varepsilon}}\ln\left(\frac{1}{\varepsilon^{1/4}S_\varepsilon}\right)T L\varepsilon S_\varepsilon P \left(1+LS_\varepsilon e^{LS_\varepsilon}\right)\nonumber\\ 
 & \hspace{1.2cm} e^{T L\big(1+ S_\varepsilon L e^{LS_\varepsilon}\big)}\nonumber\\ 
 &\quad \stackrel{\eqref{eq:se_definition}}{=} P \lim_{\varepsilon\to0}\sqrt\varepsilon S_\varepsilon \left(\ln\left(\frac{1}{\varepsilon^{1/4}S_\varepsilon}\right)\right)^2 \frac{1}{\varepsilon^{1/4}S_\varepsilon}\nonumber\\
 &\quad = P \lim_{\varepsilon\to0} \frac{1}{S_\varepsilon}\varepsilon^{1/4} S_\varepsilon \left(\ln\left(\frac{1}{\varepsilon^{1/4}S_\varepsilon}\right)\right)^2\stackrel{\eqref{eq:se}}{=}0,\label{eq:limb2}
\end{align}
where we used $\lim_{x\to0} x (\ln\frac{1}{x})^2 = 0$.
\begin{align}
 %----- 3
(iii):\quad &\lim_{\varepsilon\to0}\frac{1}{\sqrt{\varepsilon}}\ln\left(\frac{1}{\varepsilon^{1/4}S_\varepsilon}\right)T L\varepsilon \left(1+LS_\varepsilon e^{LS_\varepsilon}\right)\nonumber\\ 
 & \hspace{1cm} e^{T L\big(1+ S_\varepsilon L e^{LS_\varepsilon}\big)}\nonumber\\ 
 &\quad \stackrel{\eqref{eq:se_definition}}{=} \lim_{\varepsilon\to0} \sqrt\varepsilon \left(\ln\left(\frac{1}{\varepsilon^{1/4}S_\varepsilon}\right)\right)^2 \frac{1}{\varepsilon^{1/4}S_\varepsilon}\nonumber\\
 &\quad = \lim_{\varepsilon\to0} \frac{1}{(S_\varepsilon)^2}\varepsilon^{1/4} S_\varepsilon \left(\ln\left(\frac{1}{\varepsilon^{1/4}S_\varepsilon}\right)\right)^2\stackrel{\eqref{eq:se}}{=}0.\label{eq:limb3}
\end{align}
The proof of Lemma~\ref{lemma:DDelta} is now complete.
%========================================================
\bibliographystyle{IEEEtran}
\bibliography{reference}
%========================================================

\end{document}